\documentclass[%
 reprint,
 amsmath,amssymb,
 aps,
]{revtex4-2}

\usepackage{graphicx}
\usepackage{dcolumn}
\usepackage{bm}
\usepackage{hyperref}
\usepackage{slashed}
\usepackage{subfloat}
\usepackage{pgfplots}
\usepackage{subfigure}
\usepackage{etoolbox}   
\usepackage{orcidlink}
\usepackage{geometry}
\definecolor{acsblue}{RGB}{17,76,139}

\begin{document}

\fontsize{7.6}{8.6}\selectfont
\preprint{APS/123-QED}

\title{Lorentz-Violating Wormhole Optics}

\author{Omar Mustafa
}
\email{omar.mustafa@emu.edu.tr}
\affiliation{Department of Physics, Eastern Mediterranean University, 99628, G. Magusa, north Cyprus, Mersin 10 - Türkiye}

\author{Semra Gurtas Dogan
}
\email{semragurtasdogan@hakkari.edu.tr}
\affiliation{Department of Medical Imaging Techniques, Hakkari University, 30000, Hakkari, Türkiye}

\author{Abdulkerim Karabulut
}
\email{akerimkara@gmail.com}
\affiliation{Department of Basic Sciences, Erzurum Technical University, 25050, Erzurum, Türkiye}

\author{Abdullah Guvendi
}
\email{abdullah.guvendi@erzurum.edu.tr (Corresponding Author) }
\affiliation{Department of Basic Sciences, Erzurum Technical University, 25050, Erzurum, Türkiye}

\date{\today}

\begin{abstract}
{\fontsize{7.6}{8.6}\selectfont \setlength{\parindent}{0pt}
We study massless spin-1 field propagation in a static, circularly symmetric $(2+1)$-dimensional wormhole with spatial Lorentz-violating anisotropy characterized by the throat radius $a$ and deformation parameter $\eta$. The geometry is horizon-free, geodesically complete, and asymptotically flat, with negative Gaussian curvature localized near the throat. Using the fully covariant vector boson formalism and covariant Maxwell theory, we derive an exact Schr\"odinger-type radial equation with a curvature-induced effective potential. Recasting the dynamics in Helmholtz form yields an effective refractive-index profile, showing that the wormhole acts as an inhomogeneous optical medium with position-dependent refractive index and frequency-dependent confinement, where low-frequency modes are strongly trapped while high-frequency modes propagate almost freely. A differential-geometric correspondence with helicoidal surfaces is established via $1/[a^2(1-\eta)] \leftrightarrow w^2$, demonstrating that Lorentz-violation-induced curvature is mathematically equivalent to curvature generated by geometric twist and linking the model to twisted graphene nanoribbons as analog-gravity platforms. These results provide a geometric framework for curvature-driven localization, dispersion, and anisotropic wave propagation in topologically nontrivial $(2+1)$-dimensional backgrounds.
}
\end{abstract}


\maketitle

\tableofcontents

\section{Introduction}

\setlength{\parindent}{0pt}

Vector bosons represent the gauge fields that mediate fundamental interactions in relativistic quantum field theory and emerge naturally from Abelian and non-Abelian gauge symmetries \cite{Griffiths}. Depending on the symmetry structure and spontaneous symmetry-breaking mechanism, gauge bosons may remain massless, such as the photon associated with an unbroken $U(1)$ gauge symmetry, or acquire mass through the Higgs mechanism, as in the case of the electroweak $W^\pm$ and $Z^0$ bosons \cite{Griffiths,peskin}. The Higgs mechanism generates longitudinal polarization states and introduces a finite interaction range, whereas massless vector bosons possess only transverse polarizations and mediate long-range forces. In curved spacetime, vector boson dynamics are governed by fully covariant field equations that generalize Maxwell and Proca theories, with curvature modifying dispersion relations, polarization modes, and spatial localization of fields \cite{jackson,vb-1,vb-2}. In the frequency domain, Maxwell’s equations reduce to the Helmholtz equation, corresponding to the gauge-invariant massless limit of the Proca equation, revealing a direct correspondence between classical wave propagation and relativistic vector boson theory \cite{jackson}. Analytical solutions of vector boson equations in curved backgrounds are therefore essential for quantifying curvature-induced quantum effects \cite{jackson,vb-1,vb-2} and understanding how spacetime geometry influences relativistic quantum dynamics \cite{vb-3,cavit,thorne,1,2,3,4,5,6,wormhole,epjc,7,8,9,9.1,PM}.

\vspace{0.15cm}

Modified gravity theories provide frameworks in which nontrivial spacetime topologies, including wormholes \cite{TW1,TW2,TW3,TW4,Morris,TW5,Falco1,Falco2,Falco3}, can exist without requiring exotic matter that violates classical energy conditions. In these scenarios, apparent violations of energy conditions arise from geometric corrections, higher-order curvature terms, or background tensor fields rather than from physical matter sources \cite{TW6,TW7,TW8,TW9,Radhakrishnan2024}. Lorentz invariance is a foundational symmetry in both general relativity and the Standard Model, guaranteeing the universality of the speed of light and defining the causal structure of spacetime \cite{Kibble,Smolin,C-1,C-2}. Nevertheless, Lorentz symmetry is expected to be approximate, potentially broken in quantum gravity or high-energy effective field theories \cite{Kibble,Smolin,C-1,C-2,C-3,C-4,C-5}. The Standard-Model Extension (SME) constitutes a general effective field-theory framework for describing Lorentz and charge conjugation-parity-time reversal (CPT) violation through background tensor fields acquiring nonzero vacuum expectation values, thereby introducing preferred spacetime directions while maintaining observer covariance \cite{KS1989,ColladayKostelecky1997,C-2,C-1}. Numerous studies have placed stringent bounds on Lorentz- and CPT-violating coefficients across multiple sectors, including CPT tests \cite{23,24,25,26}, radiative corrections \cite{27,28,29,30,31,32,33,34}, gauge-field sectors \cite{35,36,37,38,39}, photon-fermion interactions \cite{40,41,42,43}, and fermionic couplings \cite{44,45,46,47}.

\vspace{0.15cm}

In gravitational models with spontaneous Lorentz symmetry breaking, wormhole geometries can arise without exotic matter. Einstein-bumblebee gravity \cite{bumblebee,Eslam,Ovgun2019,Oliveira2018,Ding2024,Maluf2020,Ghosh2024,MalufMuniz2021,KS1989,C-1,LV} offers a prominent example in which a vector field acquires a vacuum expectation value, generating Lorentz violation and supporting traversable wormhole solutions whose throat structure is controlled by Lorentz-violating coefficients \cite{Ovgun2019}. Perturbative stability analyses and quasi-normal mode studies indicate consistency within the effective-field-theory regime \cite{Oliveira2018}. Extensions that include phantom scalar fields enlarge the class of Lorentz-violating wormhole solutions and show that Lorentz symmetry breaking alone can sustain nontrivial spacetime topology \cite{Ding2024}. Additional Lorentz-violating backgrounds, such as antisymmetric Kalb-Ramond tensor fields motivated by low-energy string theory, modify black hole and wormhole metrics and influence horizons, thermodynamics, and particle trajectories \cite{Maluf2020,Ghosh2024,MalufMuniz2021}. Thin-shell constructions further connect Lorentz-violating black holes to traversable wormholes, reinforcing the geometric origin of exotic topologies in Lorentz-violating gravity \cite{KS1989,C-1,LV}.

\vspace{0.15cm}

Lower-dimensional spacetimes provide analytically tractable models for studying quantum field propagation in curved geometries, where curvature-induced potentials, bound states, and topological effects can be derived explicitly \cite{Guvendi-PLB-1,Guvendi-PLB-2}. In Lorentz-violating wormhole backgrounds \cite{LV,LV-1}, curvature generates effective potentials that modify vector bosonic modes, produce discrete spectra, and alter vacuum fluctuations near the wormhole throat \cite{abdelghani-2025,o1}. In this work, we analyze relativistic spin-1 vector bosons propagating in a $(2+1)$-dimensional Lorentz-violating wormhole geometry using the fully covariant vector boson formalism \cite{vb-1,vb-2,vb-3}. We derive the exact wave equations, construct curvature-induced effective potentials, and demonstrate that the Lorentz-violating parameter $\eta$\footnote{The parameter \(\eta\) represents an effective Lorentz-violating anisotropy in the reduced spatial sector rather than a full SME background field.} controls curvature localization, geometric confinement, and mode dispersion. By mapping the wave equation to a Helmholtz form with an effective refractive index, we show how curvature and Lorentz symmetry breaking modify photonic propagation in a topologically nontrivial background. A geometric correspondence is also established between Lorentz-violating wormholes \cite{LV,LV-1} and helicoidal (twisted) surfaces \cite{reff}, showing that Lorentz-violation-induced curvature profiles are mathematically equivalent to those generated by finite geometric twist. This correspondence connects gravitational Lorentz violation with experimentally realizable analog systems such as twisted graphene nanoribbons, enabling laboratory investigations of curvature-driven wave dynamics. 

\vspace{0.15cm}

The paper is organized as follows. In Sec.~\ref{sec:2}, we introduce the three dimensional (3D) Lorentz-violating wormhole geometry and analyze its curvature and embedding properties. In Sec.~\ref{sec:3}, we derive the covariant vector boson wave equation, construct the effective curvature-induced potential, analyze photonic mode propagation, and establish the helicoidal analog interpretation. Section~\ref{sec:4} summarizes the main results and discusses implications for Lorentz-violating gravity, analog gravity systems, and curvature-driven quantum field dynamics in nontrivial topologies.

\begin{figure*}[!ht]
\centering
\includegraphics[width=0.95\linewidth]{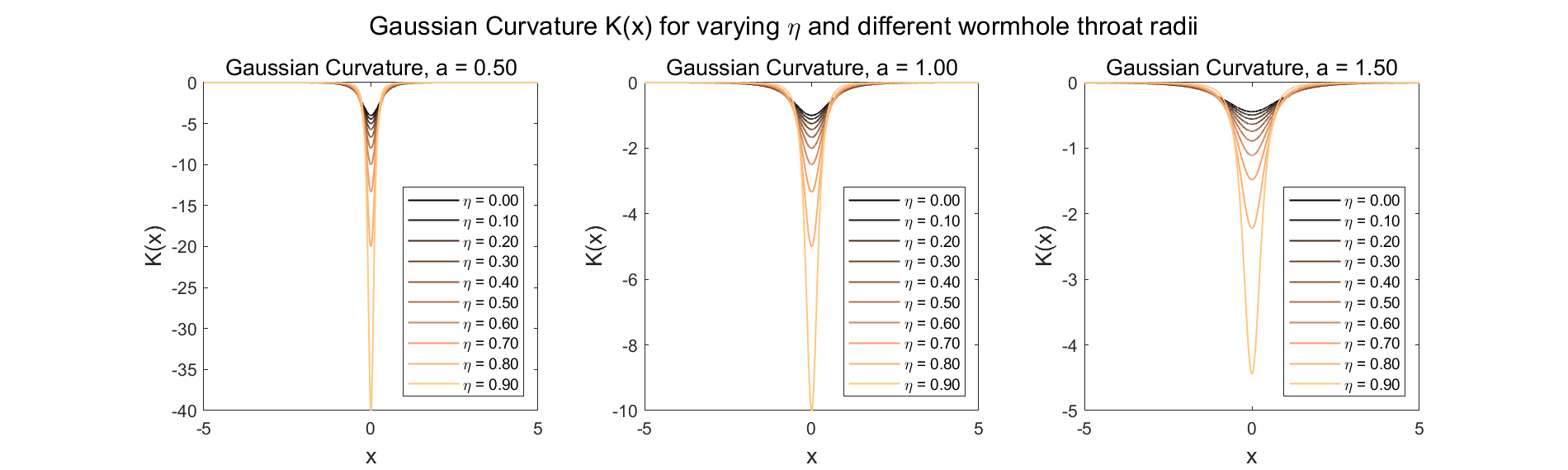}
\caption{\fontsize{7.6}{8.6}\selectfont Gaussian curvature $\mathcal{K}(x)$ of a $(2+1)$-dimensional Lorentz-violating wormhole for varying Lorentz-violation parameters $\eta \in [0,0.9]$ and three different throat radii $a = 0.5, 1.0, 1.5$. Each subplot corresponds to a fixed wormhole throat, while the curves within each subplot represent increasing values of $\eta$. The curvature is strictly negative throughout, indicating a hyperbolic optical geometry, and the magnitude of $K$ near the throat increases with both decreasing throat radius and increasing Lorentz-violation parameter, reflecting the enhancement of geometric flaring due to Lorentz-symmetry breaking.}
\label{fig:Gaussian-curvature}
\end{figure*}

\section{Lorentz-violating 3D wormhole} \label{sec:2}

\setlength{\parindent}{0pt}

We consider a static and circularly symmetric $(2+1)$-dimensional wormhole spacetime \cite{Guvendi-PLB-1} incorporating an effective Lorentz-violating deformation in the spatial sector \cite{LV}. Lower-dimensional wormhole geometries provide an analytically tractable framework for exploring quantum field propagation, curvature-induced quantum states, and topological effects in curved spacetime, while Lorentz-violating deformations allow for controlled departures from isotropic relativistic symmetry that can arise in effective field theories, analog gravity systems, and condensed-matter-inspired emergent metrics. The spacetime line element can be taken as \cite{Guvendi-PLB-1,LV}
\begin{equation}
\begin{split}
ds^2 = g_{\mu\nu} dx^\mu dx^\nu 
= A(x)\, c^2\,dt^2 - A^{-1}(x)\, dx^2 - r^2(x)\, d\theta^2,\label{metric}
\end{split}
\end{equation}
where $\mu,\nu = (t,x,\theta)$, $c$ is the speed of light, $t \in (-\infty,\infty)$ is the temporal coordinate, $x \in (-\infty,\infty)$ is a proper radial coordinate spanning the wormhole throat and both asymptotic regions, $\theta \in [0,2\pi]$ parametrizes the compact angular direction. The geometry is fully characterized by the areal radius function $r(x)$, which determines the intrinsic curvature and the flaring-out behavior of the wormhole. We adopt the areal radius profile \cite{LV,LV-1}
\begin{equation}
r(x) = \sqrt{a^2 + \frac{x^2}{1-\eta}}, 
\qquad a>0, \quad 0 \le \eta < 1,
\end{equation}
where $a$ is the throat radius and $\eta$ is a dimensionless deformation parameter that rescales the radial sector relative to the angular sector. This deformation can be interpreted as an effective Lorentz-violating anisotropy in the reduced-dimensional geometry, modifying the spatial propagation structure and curvature localization. The condition $0 \le \eta < 1$ ensures that the geometry remains real, smooth, and free of signature pathologies. The redshift function is chosen as $A(x)=1$, which ensures that the geometry is globally static, horizon-free, and geodesically complete \cite{LV,LV-1}. This choice guarantees causal connectivity between the two asymptotically flat regions and avoids gravitational redshift divergences, thereby providing a clean background for analyzing quantum fields and wave dynamics. For this class of metrics, the covariant and contravariant metric tensors are \cite{Guvendi-PLB-1}
\begin{equation}
g_{\mu\nu} = \mathrm{diag}(c^2, -1, -r^2(x)), 
\qquad 
g^{\mu\nu} = \mathrm{diag}(c^{-2}, -1, -r^{-2}(x)),
\end{equation}
with determinant
\begin{equation}
g = \det(g_{\mu\nu}) = c^2r^2(x),
\end{equation}
confirming the metric signature $(+,-,-)$. The Levi-Civita connection coefficients follow directly from the metric, yielding the only non-vanishing Christoffel symbols \cite{Guvendi-PLB-1}
\begin{equation}
\Gamma^x_{\theta\theta} = - r r', 
\qquad 
\Gamma^\theta_{x\theta} = \Gamma^\theta_{\theta x} = \frac{r'}{r},
\end{equation}
where primes denote derivatives with respect to $x$. The curvature is encoded in the Riemann tensor, whose only independent non-zero component is \cite{SOA-2025}
\begin{equation}
R_{x\theta x\theta} = r r'',
\end{equation}
leading to the Ricci tensor components
\begin{equation}
R_{xx} = \frac{r''}{r}, 
\qquad 
R_{\theta\theta} = r r'',
\end{equation}
and the Ricci scalar \cite{SOA-2025}
\begin{equation}
R = g^{\mu\nu} R_{\mu\nu} 
= -2 \frac{r''}{r}.
\label{Ricci-scalar}
\end{equation}
For the two-dimensional spatial slices at constant $t$, the Gaussian curvature is related to the Ricci scalar via $\mathcal{K} = R/2$, giving \cite{Guvendi-PLB-1}
\begin{equation}
\mathcal{K}(x) = -\frac{r''}{r}.
\end{equation}
Using the explicit form of $r(x)$, one obtains (see also \cite{Guvendi-PLB-1})
\begin{equation}
\mathcal{K}(x) 
= - \frac{a^2 (1-\eta)}{\left[a^2 (1-\eta) + x^2 \right]^2}.
\label{gaussian-curvature}
\end{equation}
This expression demonstrates that the curvature is strictly negative for all $x$, implying that the spatial geometry is everywhere hyperbolic. The curvature magnitude attains its maximum at the throat,
\begin{equation}
|\mathcal{K}(0)| = \frac{1}{a^2 (1-\eta)},
\end{equation}
showing that decreasing the throat radius $a$ or increasing the deformation parameter $\eta$ enhances curvature localization at the throat. For large $|x|$, the curvature decays as
\begin{equation}
\mathcal{K}(x) \sim -\frac{a^2(1-\eta)}{x^4},
\end{equation}
confirming that the geometry is asymptotically flat and approaches a Minkowski-like region far from the throat. The Lorentz-invariant limit is obtained by setting $\eta \to 0$, in which case
\begin{equation}
r(x) = \sqrt{a^2 + x^2},
\end{equation}
corresponding to the well-known $(2+1)$-dimensional Ellis wormhole geometry. The Gaussian curvature reduces to \cite{Guvendi-PLB-1}
\begin{equation}
\mathcal{K}_{\eta=0}(x) = -\frac{a^2}{(a^2 + x^2)^2},
\end{equation}
which is the standard curvature profile of the Ellis wormhole. Therefore, the present geometry represents a continuous Lorentz-violating deformation of the Ellis wormhole, with $\eta$ parametrizing anisotropic deviations from Lorentz invariance. The spatial geometry can be visualized via an isometric embedding into Euclidean 3D space. At fixed $t$, the spatial metric is
\begin{equation}
dl^2 = dx^2 + r^2(x) d\theta^2.
\end{equation}
Embedding this surface in cylindrical coordinates $(z,r,\theta)$ yields \cite{Guvendi-PLB-1}
\begin{equation}
dl^2 = \left[1 + \left(\frac{dz}{dx}\right)^2 \right] dx^2 + r^2(x) d\theta^2,
\end{equation}
which implies
\begin{equation}
\frac{dz}{dx} = \pm \sqrt{\frac{r'^2}{1 - r'^2}},
\end{equation}
with
\begin{equation}
r'(x) = \frac{x}{(1-\eta) r(x)}.
\end{equation}

\begin{figure}[!ht]
\centering
\includegraphics[width=0.90\linewidth]{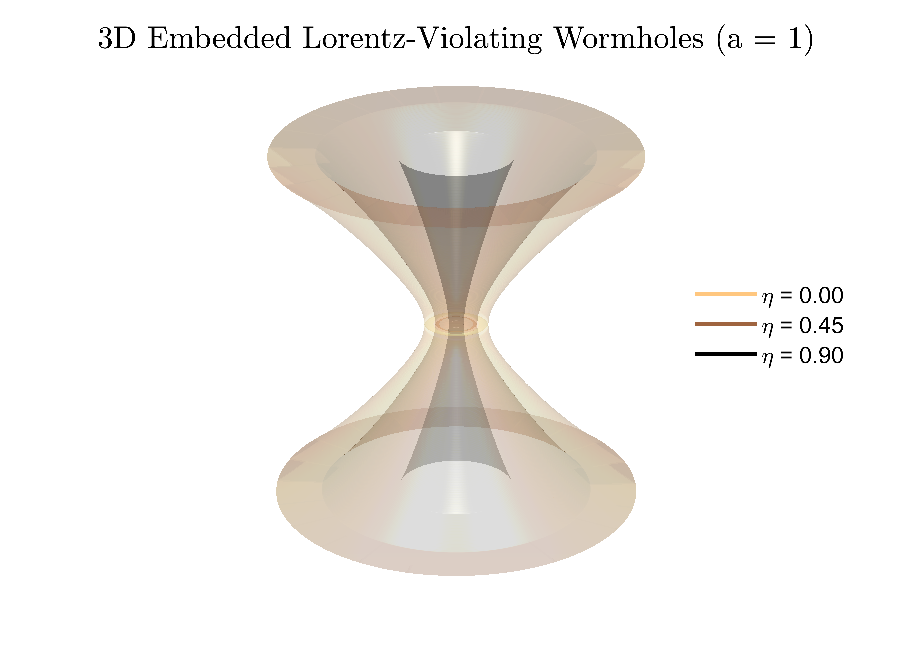}
\caption{\fontsize{7.6}{8.6}\selectfont 3D embedding diagrams of Lorentz-violating wormholes (with $x_{\text{max}}= 8$) for different values of the Lorentz-violation parameter $\eta$ at fixed throat radius $a=1$. The surfaces correspond to $\eta = 0$, $\eta = 0.45$, and $\eta = 0.9$. Both the upper ($z>0$) and lower ($z<0$) embedding branches are displayed to show the full wormhole geometry. Increasing $\eta$ significantly modifies the radial stretching of the embedding surface, indicating a strong deformation of the wormhole geometry induced by Lorentz-violating effects.}
\label{fig:3D}
\end{figure}
The embedding surface flares outward from the throat, and the flaring-out condition $r''(0)>0$ is satisfied, confirming that the geometry represents a genuine traversable wormhole rather than a coordinate artifact. The spatial topology consists of two asymptotically flat regions connected by a throat. The deformation parameter $\eta$ plays a crucial physical and geometric role. 
Geometrically, it rescales the proper radial distance relative to the angular sector, thereby controlling curvature localization, throat sharpness, and the global embedding shape. Physically, $\eta$ can be interpreted as an effective Lorentz-violating coefficient that modifies dispersion relations and propagation characteristics for fields confined to the curved surface. In analog gravity realizations, such anisotropies arise naturally due to lattice effects, strain-induced metrics, or background gauge fields. 
Increasing $\eta$ redistributes and concentrates curvature near the throat, enhancing geometric confinement and potentially supporting curvature-induced bound states and quasi-normal modes, while the limit $\eta \to 0$ restores isotropic Lorentz-invariant propagation. Fig.~\ref{fig:Gaussian-curvature} shows the Gaussian curvature profiles for various values of the deformation parameter $\eta$ and throat radius $a$. Larger values of $\eta$ produce sharper curvature localization at the throat, whereas smaller throat radii $a$ increase the curvature scale and accelerate its spatial decay. Conversely, larger $a$ smooths the curvature distribution and extends the region of geometric influence. Thus, the parameter space $(a,\eta)$ provides a tunable geometric laboratory in which curvature localization, throat geometry, and Lorentz-violating deformation can be independently controlled. Fig.~\ref{fig:3D} displays the embedded 3D geometry of Lorentz-violating wormholes for several values of $\eta$ at fixed throat radius $a=1$. As $\eta$ increases, the embedding surface becomes progressively more stretched along the radial direction, indicating that Lorentz-violating corrections effectively deform the spatial geometry and modify the throat profile. 
This behavior reflects a modification of the areal radius function and the associated shape function, leading to a widening and reshaping of the wormhole geometry in the embedding space. The presence of both positive and negative embedding branches demonstrates the symmetric nature of the wormhole spacetime, connecting two asymptotically flat regions. The Lorentz-violating wormhole introduced here is globally static, traversable, and asymptotically flat, with controllable anisotropic deformation. 
It provides a mathematically consistent and physically motivated background for investigating quantum field dynamics, vector boson propagation, wave localization, curvature-induced quantum effects, and Lorentz-symmetry breaking phenomena in $(2+1)$-dimensional curved spacetime.

\section{Photonic modes}\label{sec:3}

\setlength{\parindent}{0pt}

\begin{figure*}[!htbp]
\centering
\includegraphics[width=0.90\linewidth]{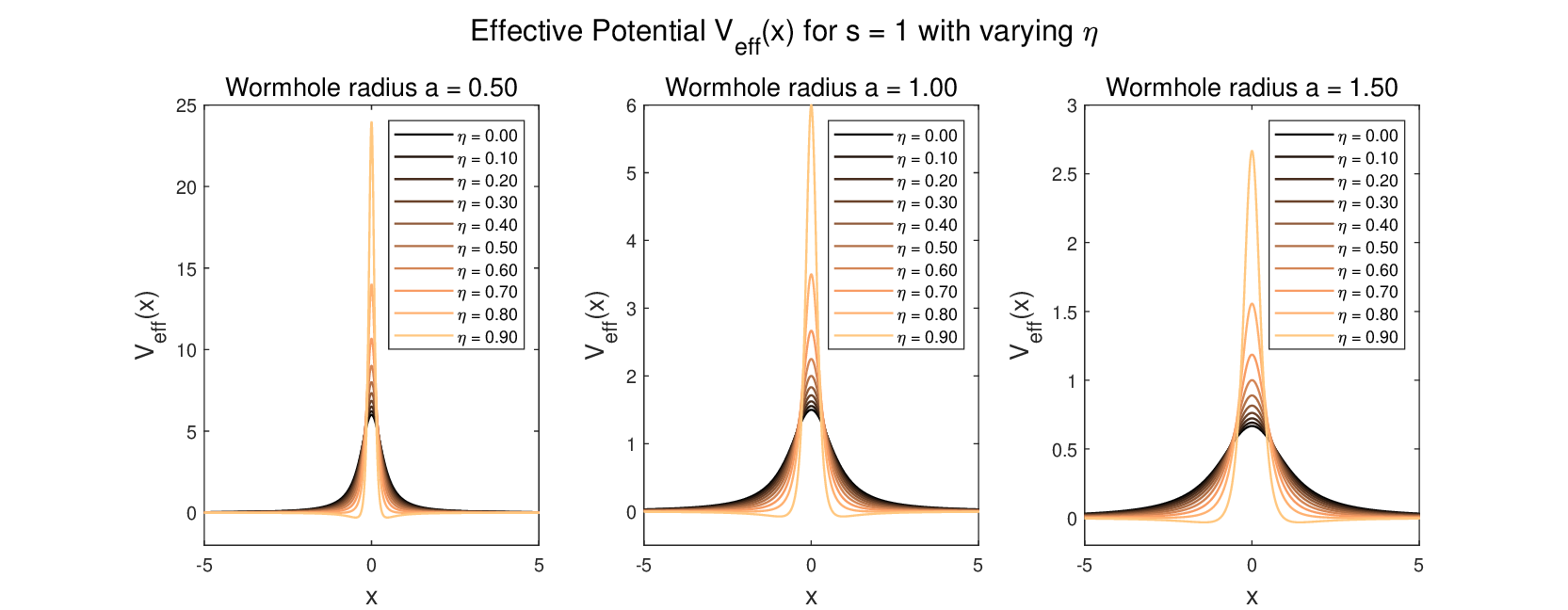}
\caption{\fontsize{8}{9}\selectfont Effective potential $V_{\rm eff}(x)$ \eqref{eq:Veff-explicit} for massless spin-1 vector bosons in the Lorentz-violating wormhole. The potential is plotted for $a=0.5, 1.0, 1.5$ and $\eta \in [0,0.9]$, showing the deformation induced by Lorentz violation.}
\label{fig:eff-pot}
\end{figure*}

We investigate the propagation of relativistic spin-1 bosonic excitations in the Lorentz-violating wormhole geometry introduced in Sec.~\ref{sec:2}. The analysis is performed using the fully covariant vector boson formalism in $2+1$ dimensions, corresponding to the spin-1 sector of the Duffin-Kemmer-Petiau equation \cite{vb-1,vb-2,vb-3}. This formalism provides a first-order relativistic description of vector bosons, allowing an exact treatment of the effects of curvature and Lorentz violation on massless and massive vector bosonic fields. The goal is to derive the complete radial wave equation governing vector boson propagation, construct the effective geometric potential, and analyze the structure and localization properties of the resulting photonic modes. In $(2+1)$-dimensional curved spacetime, the generalized vector boson equation can be written as \cite{vb-1,vb-2,vb-3}
\begin{equation}
\left[ \mathcal{B}^{\mu} \slashed{\nabla}_{\mu} + i \tilde{m} \mathbf{I}_4 \right] \Psi(x^\mu) = 0,
\label{eq:vector-boson}
\end{equation}
where $\tilde{m}=m_b\,c/\hbar$, $m$ is the rest mass of the boson, $\hbar$ is the reduced Planck constant, $\Psi(x^\mu)$ is the symmetric bispinor field, and $\mathcal{B}^\mu$ are the generalized spin-1 matrices constructed from the curved-space Dirac matrices $\gamma^\mu$ as \cite{vb-1,vb-2,vb-3}
\begin{equation}
\mathcal{B}^\mu = \frac{1}{2}\left( \gamma^\mu \otimes \mathbf{I}_2 + \mathbf{I}_2 \otimes \gamma^\mu \right),
\end{equation}
with $\otimes$ denoting the Kronecker product and $\mathbf{I}_2$ the $2 \times 2$ identity matrix. The covariant derivative acting on $\Psi$ is defined by \cite{vb-1,vb-2,vb-3}
\begin{equation}
\slashed{\nabla}_\mu = \partial_\mu - \Omega_\mu, \qquad \Omega_\mu = \Gamma_\mu \otimes \mathbf{I}_2 + \mathbf{I}_2 \otimes \Gamma_\mu,
\end{equation}
where $\Gamma_\mu$ are the usual spin connections for Dirac fields. These connections encode the geometric structure of spacetime and ensure local Lorentz covariance.

\begin{figure*}[!ht]
\centering
\includegraphics[width=1\linewidth]{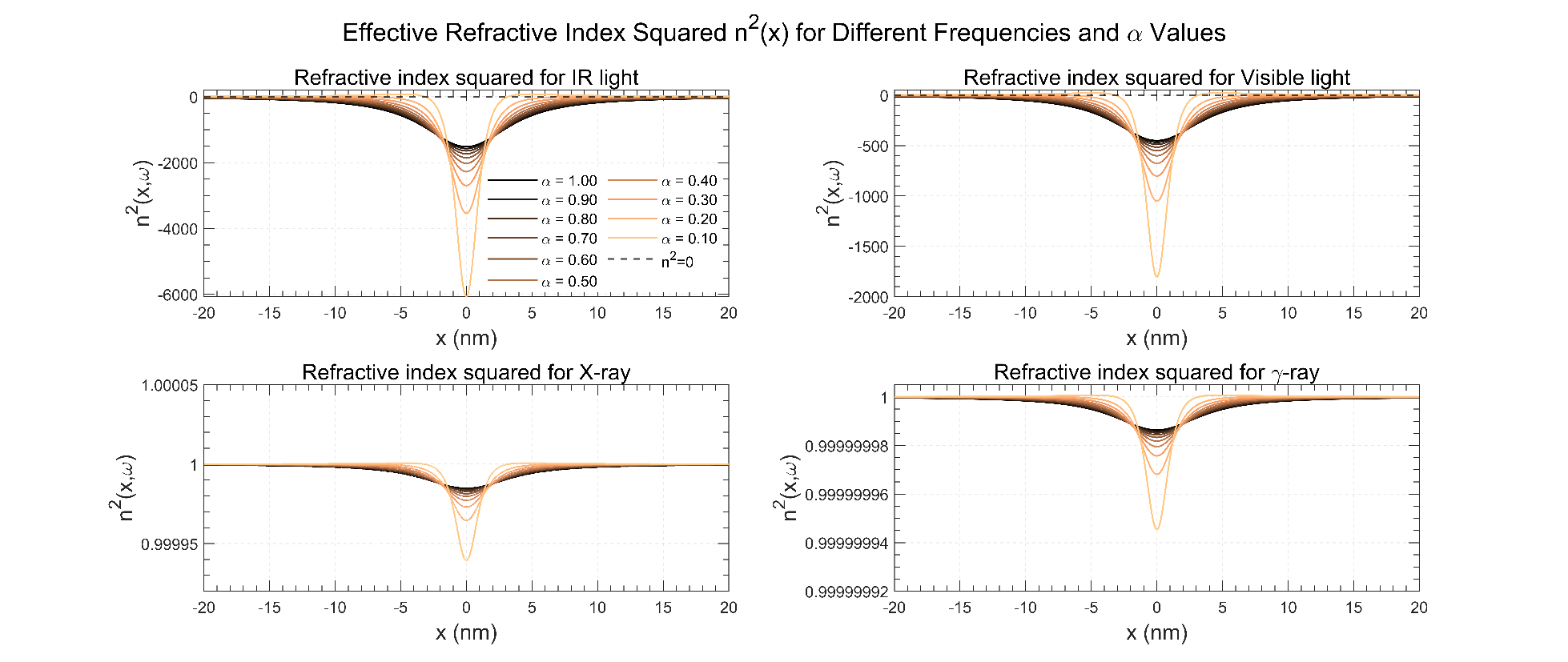}
\caption{\fontsize{8}{9}\selectfont Refractive index squared $n^2(x)$ \eqref{R-Index} for massless spin-1 vector boson propagation in the Lorentz-violating wormhole geometry for four representative electromagnetic frequencies: IR ($\omega = 3 \times 10^{14}$~Hz), visible light ($\omega = 5.5 \times 10^{14}$~Hz), X-ray ($\omega = 3 \times 10^{18}$~Hz), and gamma-ray ($\omega = 1 \times 10^{20}$~Hz). The wormhole throat radius is taken as $a = 5$~nm, and the spin of the propagating mode is $s=1$. The curves correspond to ten values of the Lorentz-violation parameter $\eta$ in the range $[0,0.9]$, with $\alpha = 1-\eta$, showing how Lorentz-violation modifies the effective refractive index profile along the longitudinal coordinate $x \in [-20,20]$~nm. The black dashed line indicates $n^2(x)=0$, separating propagating regions ($n^2>0$) from evanescent or exponentially decaying regions ($n^2<0$). The spatial variations are localized near the wormhole throat, while far from the throat, $n^2(x)$ asymptotically approaches unity. }
\label{fig:ref-index}
\end{figure*}

\vspace{0.10cm}
\setlength{\parindent}{0pt}

We consider a static axially symmetric metric of the form \eqref{metric} and specialize to the Lorentz-violating wormhole configuration with a constant lapse function, \(A(x) = 1\), which ensures a globally static, horizon-free geometry. The areal radius function is \(r(x) = \sqrt{a^2 + \frac{x^2}{1-\eta}}, \quad \eta < 1\), where $a$ is the throat radius and $\eta$ is the Lorentz-violating parameter. The wormhole throat is located at $x=0$, with $r'(0)=0$ and $r''(0)>0$, guaranteeing smooth geometry and the absence of curvature singularities. Far from the throat, $|x| \to \infty$, the geometry approaches quasi-Minkowski space with negligible curvature contributions. A convenient triad representation yields the curved-space Dirac matrices \cite{vb-1,vb-2,vb-3,Guvendi-PLB-1}
\begin{equation}
\gamma^t = \frac{1}{c}\sigma^z, \qquad \gamma^x = i \sigma^x, \qquad \gamma^\theta = i r^{-1}(x) \sigma^y,
\end{equation}
where $\sigma^{x}$, $\sigma^{y}$ and $\sigma^z$ are the Pauli matrices. The only nonvanishing spin connection component is \cite{vb-1,vb-2,Guvendi-PLB-1}
\begin{equation}
\Gamma_\theta = \frac{i}{2} r'(x) \sigma^z.
\end{equation}
This component captures the intrinsic curvature of the azimuthal direction and enters the spin-1 covariant derivative through $\Omega_\theta$. We separate the vector boson field as \cite{vb-1,vb-2}
\begin{equation}
\Psi(x^\mu) = e^{-i\omega t} e^{i s \theta} 
\begin{pmatrix} \psi_1(x) \\ \psi_2(x) \\ \psi_3(x) \\ \psi_4(x) \end{pmatrix},
\end{equation}
where $\omega$ is the frequency and $s$ is the spin projection along the axis of symmetry. Introducing the combinations \cite{vb-1,vb-2}
\begin{equation}
\psi_0 = \psi_2 + \psi_3, \qquad \psi_\pm = \psi_1 \pm \psi_4,
\end{equation}
the vector boson system reduces to the three equations including coupled first-order equations \cite{AO-Int}
\begin{align}
&\tilde{\omega} \psi_+ - m_b \psi_- - \frac{s}{r} \psi_0 = 0, \label{eq:first} \\
&\tilde{\omega} \psi_- - m_b \psi_+ - \psi^{'}_0 = 0, \label{eq:second} \\
&m_b \psi_0 - \frac{s}{r} \psi_- + \left(\frac{d}{dx} + \frac{r'}{r}\right) \psi_+ = 0, \label{eq:third}
\end{align}
where $\tilde{\omega}=\omega/c$. Eliminating the components $\psi_\pm$ of the vector boson field yields a second-order differential equation for the component $\psi_0(x)$, which captures the essential dynamics in the wormhole background,
\begin{equation}
\psi^{''}_0 + \frac{r'(x)}{r(x)}  \psi^{'}_0 + \left( \lambda - \frac{s^2}{r^2(x)} \right) \psi_0(x) = 0, 
\qquad \lambda = \tilde{\omega}^2 - \tilde{m}^2.
\label{eq:radial-wave}
\end{equation}
This equation governs the propagation of vector bosons in the Lorentz-violating wormhole background. In the massless limit, $\tilde{m}^2 = 0$, it reduces to a Helmholtz equation in curved space, with $\tilde{\omega}$ acting as the wavenumber \cite{AO-Int}. To transform Eq.~\eqref{eq:radial-wave} into a Schrödinger-like form suitable for physical and optical interpretations, we perform a rescaling
\begin{equation}
\psi_0(x) = \frac{\psi(x)}{\sqrt{r(x)}},
\end{equation}
which eliminates the first derivative term and yields
\begin{equation}
\psi^{''}(x) + \left[ \lambda - V_{\rm eff}(x) \right] \psi(x) = 0,
\label{eq:schrodinger}
\end{equation}
where the effective potential $V_{\rm eff}(x)$ encodes both the centrifugal contribution from angular momentum and the geometric curvature of the wormhole,
\begin{equation}
V_{\rm eff}(x) = \frac{s^2}{r^2(x)} + \frac{r''(x)}{2 r(x)} - \frac{(r'(x))^2}{4 r^2(x)}.\label{pot}
\end{equation}
For the specific wormhole profile $r^2(x) = a^2 + x^2/\alpha$ with $\alpha = 1-\eta$, the effective potential can be written explicitly as
\begin{equation}
V_{\rm eff}(x) = \frac{s^2}{a^2 + x^2/\alpha} + \frac{2 a^2 \alpha - x^2}{4 \alpha^2 \left(a^2 + x^2/\alpha \right)^2}.
\label{eq:Veff-explicit}
\end{equation}
This potential (see Fig. \ref{fig:eff-pot}) is symmetric about the wormhole throat at $x=0$ and decays asymptotically as $V_{\rm eff}(x) \sim s^2 \alpha / x^2$ for $|x| \to \infty$, indicating that curvature-induced effects are strongly localized near the throat while centrifugal effects persist at large distances, reflecting the long-range influence of angular momentum. The Schrödinger-like form in Eq.~\eqref{eq:schrodinger} allows a direct analogy to optical wave propagation in an inhomogeneous medium. By taking $\lambda = \tilde{\omega}^2 = (\omega/c)^2$, the equation can be rewritten in the canonical Helmholtz form \cite{o2,o3,o4,o5}
\begin{equation}
\psi^{''}(x) + \tilde{\omega}^2 n^2(x) \psi(x) = 0,
\end{equation}
where the effective refractive index squared is defined as \cite{o2,o3,o4,o5,o6,o7}
\begin{equation}
n^2(x) = 1 - \frac{1}{\tilde{\omega}^2} \left[ \frac{s^2}{a^2 + x^2/\alpha} + \frac{2 a^2 \alpha - x^2}{4 \alpha^2 (a^2 + x^2/\alpha)^2} \right].\label{R-Index}
\end{equation}
In this formulation, $n^2(x)$ is dimensionless, as required for consistency in the Helmholtz equation, while $\tilde{\omega}$ has units of inverse length, ensuring that the phase factor $\tilde{\omega} x$ is dimensionless. The unity term represents the asymptotic vacuum refractive index far from the wormhole throat, governing free-wave propagation. The first subtracted term, proportional to $s^2$, corresponds to the centrifugal barrier induced by spin, reducing the local refractive index near the throat and producing a mode-dependent repulsive contribution. The second term, proportional to $(2 a^2 \alpha - x^2)$, originates purely from the curvature of the wormhole geometry. It is strongly localized near the throat and decays as $1/x^2$ for large $|x|$, producing a smooth transition to vacuum and acting as an optical potential well that can slow down and partially confine the wave.  Asymptotically, the refractive index behaves as
\begin{equation}
n^2(x) \sim 1 - \frac{s^2 \alpha}{\tilde{\omega}^2 x^2}, \quad |x| \to \infty,
\end{equation}
showing that far from the throat, the centrifugal contribution dominates while curvature effects vanish, restoring vacuum-like propagation. This mapping establishes a direct optical analogy: the wormhole geometry is equivalent to a graded-index medium where the refractive index decreases locally near the throat due to curvature and spin, producing effective slowing, potential confinement, and mode-selective propagation. Each term in $n^2(x)$ is dimensionally consistent, with $x$ and $a$ in units of length, $s$ dimensionless, and $\tilde{\omega}$ in inverse length, preserving the canonical form of the Helmholtz equation. Consequently, the effective refractive index $n(x)$ captures all mechanisms governing wave propagation in the curved geometry, including geometric confinement, angular momentum-induced mode dispersion, and asymptotic recovery to free space, providing a comprehensive, physically transparent framework for analyzing optical analogs of wormhole-like structures. This formulation can be implemented in graded-index media or metamaterials, where the effects of geometry on wave propagation may be directly observed, providing a clear description of curvature-induced modulation in topologically nontrivial structures \cite{TO1,TO2,TO3,TO4}.

\vspace{0.10cm}
\setlength{\parindent}{0pt}

The Fig. \ref{fig:ref-index} shows the plots of the effective refractive index squared, $n^2(x)$, providing a detailed visualization of how wormhole geometry, and Lorentz-violation parameter $\eta$ influence wave propagation. Near the wormhole throat ($x \approx 0$), the curvature-induced contribution to the effective potential generates a local reduction in $n^2(x)$, forming an optical potential well that can slow or confine propagating modes. The centrifugal term associated with the spin $s$ introduces a repulsive contribution, shifting the minimum of $n^2(x)$ and producing mode-dependent modifications in the potential depth. As $\eta$ increases from $0$ to $0.9$ (corresponding to decreasing $\alpha = 1-\eta$), the curvature term becomes more sharply localized, leading to deeper minima in $n^2(x)$ but narrower regions where $n^2(x)<0$, indicating stronger but more spatially confined trapping. This highlights that Lorentz-violation amplifies the local modulation of the refractive index without significantly widening the evanescent region. The effect of the wave frequency is equally significant. Low-frequency modes, such as IR ($\omega = 3 \times 10^{14}$~Hz) and visible light ($\omega = 5.5 \times 10^{14}$~Hz), experience substantial deviations in $n^2(x)$, with pronounced minima and extended variation near the throat, reflecting enhanced sensitivity to both curvature and spin-induced effects. In contrast, high-frequency waves, including X-rays ($\omega = 3 \times 10^{18}$~Hz) and gamma-rays ($\omega = 1 \times 10^{20}$~Hz), exhibit minimal relative variations in $n^2(x)$ due to the $1/\tilde{\omega}^2$ scaling, maintaining nearly vacuum-like propagation except extremely close to the throat. Far from the throat, all curves asymptotically approach $n^2(x) \to 1$, confirming that curvature and Lorentz-violation effects are localized and do not affect propagation at large distances. These results reveal a frequency-dependent, spatially localized modulation of wave propagation: low-frequency modes are strongly trapped and can experience evanescent decay near the throat, while high-frequency modes propagate largely unaffected. Increasing $\eta$ intensifies the depth of the optical potential, enhancing confinement, while simultaneously narrowing the spatial extent of the trapping region. The combination of spin, geometry, frequency, and Lorentz-violation provides a precise framework for understanding wave dynamics in topologically nontrivial structures, with direct implications for optical analogs in graded-index media and metamaterials, where similar curvature-induced or Lorentz-violating effects can be experimentally simulated and probed.

\subsection{Lorentz-Violating Wormhole Geometry and Effective Helicoidal Interpretation}\label{subsec:3:1}

We consider the static circularly symmetric $(2+1)$-dimensional Lorentz-violating (LV) wormhole metric in \eqref{metric}
in which $a$ is the throat radius and $\eta$ parametrizes explicit Lorentz-symmetry breaking through an anisotropic deformation of the spatial sector. The constant-time spatial slice is
\begin{equation}
d\ell^2 = dx^2 + r^2(x) d\theta^2,
\end{equation}
The Gaussian curvature of the spatial slice is
\begin{equation}
\mathcal{K}_{\rm WH}(x)=-\frac{r''}{r}=-\frac{a^2(1-\eta)}{[a^2(1-\eta)+x^2]^2}<0,
\end{equation}
which satisfies the flaring-out condition at the throat and decays as $x^{-4}$, ensuring asymptotic flatness. The parameter $\eta$ controls the curvature concentration near the throat, with $|K(0)|=(1-\eta)/a^2$.

\vspace{0.10cm}
\setlength{\parindent}{0pt}

For comparison, a helicoidal (twisted) surface described by $\vec r(u,v)=(v,u\cos wv,u\sin wv)$ induces the metric $d\ell^2=du^2+(1+w^2u^2)dv^2$ and Gaussian curvature $\mathcal{K}_{\rm hel}(u)=-w^2/(1+w^2u^2)^2$ \cite{reff}. Both geometries share the canonical form $d\ell^2=d\rho^2+f^2(\rho)d\theta^2$ with $f_{\rm WH}^2=a^2+x^2/(1-\eta)$ and $f_{\rm hel}^2=1+w^2u^2$, implying a functional identity of curvature profiles under the correspondence
\begin{equation}
\frac{1}{a^2(1-\eta)} \leftrightarrow w^2.
\end{equation}
Therefore, the Lorentz-violation parameter $\eta$ acts as a geometric regulator that stretches the angular sector relative to the radial direction, producing a hyperbolic spatial geometry mathematically equivalent to that generated by a finite helicoidal deformation. In this precise differential-geometric sense, Lorentz violation induces an effective twist of the spatial manifold, yielding identical curvature invariants, geodesic structure, and asymptotic behavior. This establishes twisted graphene nanoribbons \cite{reff} as analogue-gravity realizations of Lorentz-violating wormhole geometries, enabling controlled exploration of curvature-driven quantum field effects in a laboratory setting, while remaining conceptually distinct from fundamental spacetime Lorentz symmetry breaking.

\subsection{Electromagnetic field dynamics} \label{sec:3:2}

\setlength{\parindent}{0pt}

The dynamics of the electromagnetic field in curved spacetime are governed by the covariant Maxwell equations \cite{jackson,Konoplya,Ahmed},
\begin{equation}
\nabla_\mu F^{\mu\nu} = 0,
\qquad 
F_{\mu\nu} = \nabla_\mu A_\nu - \nabla_\nu A_\mu ,
\label{maxwell}
\end{equation}
where $A_\mu$ is the vector potential and $\nabla_\mu$ denotes the metric-compatible covariant derivative. In curved geometry, Eq.~(\ref{maxwell}) can be written explicitly as \cite{jackson,Konoplya,Ahmed}
\begin{equation}
\frac{1}{\sqrt{g}} \partial_\mu \left( \sqrt{g}\, F^{\mu\nu} \right) = 0,
\end{equation}
with $g$ the determinant of the spacetime metric. Imposing the Lorenz gauge condition \cite{jackson},
\begin{equation}
\nabla_\mu A^\mu = 0,
\end{equation}
the Maxwell equations reduce to the covariant wave equation for the vector potential,
\begin{equation}
\nabla_\mu \nabla^\mu A^\nu - R^{\nu}_{\ \mu} A^\mu = 0,
\label{waveA}
\end{equation}
where $R^{\nu}_{\ \mu}$ is the Ricci tensor. In $(2+1)$ dimensions the curvature is fully encoded in the Ricci tensor, and therefore the wormhole geometry acts as an effective medium that modifies the propagation of electromagnetic waves. Due to the axial symmetry of the wormhole background, the electromagnetic modes can be decomposed as
\begin{equation}
A_\mu(t,x,\theta) = \mathcal{A}_\mu(x)\, e^{-i\omega t} e^{i m \theta},
\end{equation}
where $\omega$ is the mode frequency and $m$ is the azimuthal quantum number. This obeys an effective scalar wave equation \cite{KG},
\begin{equation}
\frac{1}{\sqrt{-g}} \partial_\mu \left( \sqrt{-g}\, g^{\mu\nu} \partial_\nu \Psi \right) = 0,
\label{kg}
\end{equation}
where $\Psi$ represents a generic electromagnetic mode function. Using $\sqrt{g}=c\, r(x)$ and the inverse metric components yields the radial equation
\begin{equation}
\frac{1}{r(x)} \frac{d}{dx} \left( r(x) \Psi' \right)
+ \frac{\omega^2}{c^2} \Psi
- \frac{m^2}{r^2(x)} \Psi = 0.
\label{radial1}
\end{equation}
Introducing the field redefinition
\begin{equation}
\Psi(x) = \frac{\psi(x)}{\sqrt{r(x)}},
\end{equation}
one obtains a Schrödinger-like wave equation,
\begin{equation}
\psi'' + \left[ \frac{\omega^2}{c^2} - V_{\mathrm{eff}}(x) \right] \psi = 0,
\label{schrodinger}
\end{equation}
with the effective curvature-induced potential (see also Eq. \eqref{pot})
\begin{equation}
V_{\mathrm{eff}}(x) 
= \frac{m^2}{r^2(x)} 
+ \frac{1}{2} \frac{r''(x)}{r(x)}
- \frac{1}{4} \left( \frac{r'(x)}{r(x)} \right)^2 .
\label{veff}
\end{equation}
It is important to emphasize that the Schrödinger-like radial equation derived from Maxwell theory is not specific to classical electromagnetism but emerges also for massless spin-1 vector bosons propagating in the Lorentz-violating wormhole geometry. Indeed, eliminating the components $\psi_\pm$ in the fully covariant vector boson field equations yields the second-order equation \eqref{eq:radial-wave}, which, in the massless limit $\tilde{m}^2=0$, becomes formally identical to the electromagnetic radial wave equation obtained from the covariant Maxwell equations. Both approaches lead to an identical Schrödinger-type equation with the same curvature-induced effective potential $V_{\rm eff}(x)$. For axisymmetric modes, the azimuthal quantum number $m$ plays a role analogous to the spin projection $s$ in the vector boson formalism, and we adopt the identification $m \equiv s$ for comparison. This demonstrates that the wormhole geometry acts as a universal geometric potential for massless spin-1 fields, independent of whether they are treated as classical electromagnetic waves or quantized vector bosons. The equivalence establishes a deep geometric origin of photonic and vector-bosonic mode confinement, showing that curvature and Lorentz-violating deformation encode an effective optical medium whose trapping and dispersion properties are fundamentally dictated by spacetime topology rather than microscopic field-theoretic details. Consequently, electromagnetic waves and massless vector bosons experience identical geometric confinement, mode dispersion, and effective refractive-index modulation, providing a framework for optical analogs and quantum field propagation in topologically nontrivial curved backgrounds. 

\section{Summary and Discussion}\label{sec:4}

\setlength{\parindent}{0pt}

In this work, we investigated the propagation of massless spin-1 vector bosons in a static, circularly symmetric $(2+1)$-dimensional Lorentz-violating wormhole geometry characterized by the throat radius $a$ and the anisotropic deformation parameter $\eta$, which explicitly breaks Lorentz symmetry in the spatial sector. The wormhole is described by the areal radius profile $r(x)=\sqrt{a^2+x^2/(1-\eta)}$, which ensures a smooth, horizon-free geometry connecting two asymptotically flat regions. As shown in Fig.~\ref{fig:Gaussian-curvature}, the Gaussian curvature $\mathcal{K}(x)=-a^2(1-\eta)/[a^2(1-\eta)+x^2]^2$ is strictly negative, confirming that the spatial slices are hyperbolic everywhere. The curvature reaches its maximum magnitude at the throat, $|\mathcal{K}(0)|=1/[a^2(1-\eta)]$, and decays as $x^{-4}$ for large $|x|$, ensuring asymptotic flatness. Increasing $\eta$ enhances curvature localization near the throat, whereas larger $a$ smooths the curvature profile and broadens its spatial influence.

\vspace{0.10cm}
\setlength{\parindent}{0pt}

The 3D embedding diagrams in Fig.~\ref{fig:3D} show the geometric deformation induced by Lorentz violation. Both upper and lower embedding branches reveal the symmetric connection between the two asymptotic regions. As $\eta$ increases, the embedding surface becomes increasingly stretched along the radial direction, indicating a stronger anisotropic deformation and sharper curvature concentration at the throat. These geometric features directly affect wave dynamics, as the curvature distribution and anisotropy determine the effective optical potential and refractive index experienced by propagating spin-1 modes.

\vspace{0.10cm}
\setlength{\parindent}{0pt}

The radial dynamics of vector bosons are governed by the Schrödinger-like equation~\eqref{eq:schrodinger} with the effective potential $V_{\rm eff}(x)$ given in Eq.~\eqref{eq:Veff-explicit} and plotted in Fig.~\ref{fig:eff-pot}. The potential consists of a centrifugal barrier associated with the spin $s$ and a curvature-induced geometric contribution, both yielding a positive and repulsive barrier localized near the throat. It is symmetric about the throat and exhibits a localized maximum determined by the combined influence between curvature and spin-induced repulsion. Increasing $\eta$ enhances the barrier height while reducing its spatial width, showing that Lorentz violation strengthens geometric repulsion while localizing it more strongly. The throat radius $a$ further controls the barrier structure: smaller $a$ increases the curvature scale and raises the barrier height, whereas larger $a$ produces a broader and weaker repulsive region.

\vspace{0.10cm}
\setlength{\parindent}{0pt}

A direct optical interpretation is provided by the effective refractive index squared $n^2(x)$ shown in Fig.~\ref{fig:ref-index}. Near the throat, low-frequency modes (infrared and visible light) exhibit pronounced reductions in $n^2(x)$, forming a localized optical potential well that can slow down or partially trap electromagnetic waves. In contrast, high-frequency modes (X-ray and gamma-ray) remain largely insensitive to curvature effects, with $n^2(x)\approx 1$ away from the throat, indicating vacuum-like propagation. Increasing $\eta$ sharpens the minima in $n^2(x)$ and narrows the region where $n^2(x)<0$, showing that Lorentz violation amplifies confinement while restricting the spatial extent of evanescent regions. The spin parameter $s$ introduces a centrifugal contribution that shifts the position and depth of the refractive-index minimum, producing mode-dependent propagation and dispersion characteristics.

\vspace{0.10cm}
\setlength{\parindent}{0pt}

A geometric correspondence between the Lorentz-violating wormhole and a helicoidal surface was established in Sec.~\ref{subsec:3:1}. The mapping $1/[a^2(1-\eta)] \leftrightarrow w^2$, where $a$ is the wormhole throat radius, $\eta$ quantifies anisotropic Lorentz-symmetry breaking, and $w=2\pi m/L$ is the helical twist density of a nanoribbon of total length $L$ with $m$ full twists \cite{reff}, shows that the Lorentz-violation parameter plays the role of an effective geometric twist. In the wormhole geometry, the deformation of the radial metric sector rescales the curvature scale as $\mathcal{K} \sim 1/[a^2(1-\eta)]$, enhancing the geometric contribution to the wave equation and strengthening the repulsive effective potential barrier for vector bosons. Under the above correspondence, this behavior is mathematically equivalent to increasing the twist density in a helicoidal nanostructure, where $w^2$ enters the wave equation as a geometric gauge field that induces centrifugal-like terms and modifies the local refractive index profile. In a differential-geometric framework, Lorentz violation induces an effective hyperbolic spatial curvature that is mathematically equivalent to the geometry generated by a helicoidally twisted surface, thereby establishing a direct correspondence between Lorentz-violating wormhole spacetimes and condensed-matter systems such as twisted graphene nanoribbons \cite{reff}. Within this correspondence, the wormhole throat radius \(a\) emerges as an effective minimal curvature length scale determined by the equivalence between the twist parameter \(w\) and the Lorentz-violation parameter \(\eta\), implying that both parameters effectively produce identical geometric deformations at the throat. This mapping furnishes a physically realizable analog platform for investigating curvature-induced wave modulation, reflection, and topological phenomena in analog gravity and engineered metamaterial systems \cite{TO1,TO2,TO3,TO4}.

\vspace{0.10cm}
\setlength{\parindent}{0pt}

The present study demonstrates that Lorentz-violating wormholes provide a tunable geometric background in which curvature, spin, frequency, and anisotropic deformation jointly determine wave propagation. Figures~\ref{fig:Gaussian-curvature} and \ref{fig:3D} quantify curvature localization and geometric deformation, Fig.~\ref{fig:eff-pot} shows the curvature and spin contributions to the effective potential, and Fig.~\ref{fig:ref-index} reveals the resulting frequency- and spin-dependent refractive-index modulation. Low-frequency modes are strongly affected by curvature-induced trapping, whereas high-frequency modes propagate almost freely. Lorentz violation concentrates curvature at the throat, deepens the effective potential, and enhances localized trapping without significantly extending the evanescent region. The helicoidal interpretation connects the gravitational model to experimentally accessible analog systems, providing a controllable platform for exploring curvature-induced bound states, quasi-normal modes, and Lorentz-symmetry-breaking wave dynamics in $(2+1)$-dimensional curved geometries. This framework bridges Lorentz-violating gravitational models with optical and condensed-matter analogs, establishing a geometric laboratory for studying wave propagation in topologically nontrivial curved spacetimes.


\section*{ Credit authorship contribution statement}

\textbf{Omar Mustafa}: Conceptualization, Methodology, Formal Analysis,  Writing - Original Draft, Investigation, Writing - Review and Editing.\\
\textbf{Semra Gurtas Dogan}: Conceptualization, Methodology, Formal Analysis, Writing - Original Draft, Investigation, Visualization, Writing - Review and Editing.\\
\textbf{Abdulkerim Karabulut}: Conceptualization, Methodology, Formal Analysis, Writing - Original Draft, Investigation, Visualization, Writing - Review and Editing.\\
\textbf{Abdullah Guvendi}: Conceptualization, Methodology, Formal Analysis, Writing - Original Draft, Investigation, Visualization, Writing - Review and Editing.\\

\section*{Data availability}

This manuscript has no associated data.

\section*{Conflicts of interest statement}

No conflict of interest declared by the authors.

\section*{Funding}

No fund has received for this study.


\nocite{*}
\begin{thebibliography}{99}
{\footnotesize
\bibitem{Griffiths} D. Griffiths, "Introduction to Elementary Particles" \href{https://doi.org/10.1002/9783527618460}{John Wiley and Sons Inc  (1987) 392 p}

\bibitem{peskin} M.E. Peskin, "An Introduction To Quantum Field Theory" \href{https://doi.org/10.1201/9780429503559}{CRC Press, Boca Raton  (1995) 866 p}

\bibitem{jackson} J.D. Jackson, "Classical Electrodynamics, 3rd Edition" \href{https://www.wiley.com/en-us/Classical+Electrodynamics%2C+3rd+Edition-p-9780471309321}{John Wiley and Sons  (1998) 832 p}

\bibitem{vb-1} A. Guvendi, S. Gurtas Dogan, "Vector boson oscillator in the near-horizon of the BTZ black hole" \href{https://doi.org/10.1088/1361-6382/acabf8}{Classical and Quantum Gravity,  \textbf{40} (2022) 025003}

\bibitem{vb-2} A. Guvendi, S. Gurtas Dogan, "Vector bosons in the rotating frame of negative curvature wormholes" \href{https://doi.org/10.1007/s10714-024-03213-z}{General Relativity and Gravitation,  \textbf{56} (2024) 32}

\bibitem{thorne} C.W. Misner, K.S. Thorne, J. A. Wheeler, "Gravitation" \href{}{Princeton University Press,  (2017) 1279 p}

\bibitem{1} L. Parker, "One-electron atom in curved space-time" \href{https://doi.org/10.1103/PhysRevD.80.124008}{Physical Review Letters, \textbf{44} (1980) 1559–1562}

\bibitem{2} L. Parker, "One-electron atom as a probe of spacetime curvature" \href{https://doi.org/10.1103/PhysRevD.22.1922}{Physical Review D, \textbf{22} (1980) 1922–1934}

\bibitem{3} A. Ashtekar, A. Magnon, "Quantum fields in curved space-times" \href{https://doi.org/10.1098/rspa.1975.0181}{Proceedings of the Royal Society A, \textbf{346} (1975) 375}

\bibitem{4} A.O. Barut, I.H. Duru, "Exact solutions of the Dirac equation in spatially flat Robertson-Walker space-times" \href{https://doi.org/10.1103/PhysRevD.36.3705}{Physical Review D, \textbf{36} (1987) 3705}

\bibitem{5} A. Guvendi, Y. Sucu, "An interacting fermion-antifermion pair in the spacetime background generated by static cosmic string" \href{https://doi.org/10.1016/j.physletb.2020.135960}{Physics Letters B, \textbf{811} (2020) 135960}

\bibitem{6} A. Guvendi, H. Hassanabadi, "Fermion-antifermion pair in magnetized optical wormhole background" \href{https://doi.org/10.1016/j.physletb.2023.138045}{Physics Letters B, 
 \textbf{843} (2023) 138045}

\bibitem{wormhole} A. Guvendi, O. Mustafa, S. Gurtas Dogan, "Coupled fermion-antifermions pairs within a traversable wormhole" \href{https://doi.org/10.1016/j.physletb.2025.139313}{Physics Letters B, 
 \textbf{862} (2025) 139313}

\bibitem{epjc} O. Mustafa, A. Guvendi, "On the Klein–Gordon scalar field oscillators in a spacetime with spiral-like dislocations in external magnetic fields" \href{https://doi.org/10.1140/epjc/s10052-025-13779-w}{European Physical Journal C, \textbf{85} (2025) 34}

\bibitem{7} P. Sedaghatnia, H. Hassanabadi, F. Ahmed, "Dirac fermions in Som–Raychaudhuri space-time with scalar and vector potential and the energy momentum distributions" \href{https://doi.org/10.1140/epjc/s10052-019-7051-6}{European Physical Journal C, \textbf{79} (2019 541)} 

\bibitem{8} G.Q. Garcia, J.R. de S. Oliveira, C. Furtado, "Weyl fermions in a family of G{\"o}del-type geometries with a topological defect" \href{https://doi.org/10.1142/S021827181850027X}{International Journal of Modern Physics D, \textbf{27} (2018) 1850027}

\bibitem{9} A. Guvendi, "Evolution of an interacting fermion--antifermion pair in the near-horizon of the BTZ black hole" \href{https://doi.org/10.1140/epjc/s10052-024-12542-x}{European Physical Journal C, \textbf{84} (2024) 1--7}.

\bibitem{9.1} 
E. Battista, H. C. Steinacker, "Fermions on curved backgrounds of matrix models"  \href{https://doi.org/10.1103/PhysRevD.107.046021}{Phys. Rev. D 107 (2023)  046021.} 

\bibitem{PM} A. Guvendi, S. Gurtas Dogan, O. Mustafa, K. Hasanirokh, "Photonic modes in twisted graphene nanoribbons" \href{https://doi.org/10.1016/j.physe.2024.116146}{Physica E: Low-dimensional Systems and Nanostructures, \textbf{166} (2025) 116146}

\bibitem{vb-3} A.O. Barut, "Excited states of zitterbewegung" \href{https://doi.org/10.1016/0370-2693(90)91202-M}{Physics Letters B, \textbf{237} (1990) 436--439}

\bibitem{cavit} Y. Sucu, C. Tekincay, "Photon in the Earth-ionosphere cavity: Schumann resonances" \href{https://doi.org/10.1007/s10509-019-3547-7}{Astrophysics and Space Science, \textbf{364} (2019) 56}

\bibitem{TW1}
M. Visser, \textit{Traversable wormholes: Some simple examples},
\href{https://doi.org/10.1103/PhysRevD.39.3182}{Phys. Rev. D, \textbf{39} (1989) 3182}.

\bibitem{TW2}
R. A. Konoplya, A. Zhidenko, \textit{Traversable wormholes in general relativity},
\href{https://doi.org/10.1103/PhysRevLett.128.091104}{Phys. Rev. Lett., \textbf{128} (2022) 091104}.

\bibitem{TW3}
J. Maldacena, A. Milekhin, F. Popov, \textit{Traversable wormholes in four dimensions},
\href{https://doi.org/10.1088/1361-6382/acde30}{Class. Quantum Grav., \textbf{40} (2023) 155016}.

\bibitem{TW4}
J. Maldacena, D. Stanford, Z. Yang, \textit{Diving into traversable wormholes},
\href{https://doi.org/10.1002/prop.201700034}{Fortschr. Phys., \textbf{65} (2017) 1700034}.

\bibitem{Morris}
M. S. Morris, K. S. Thorne,
\textit{Wormholes in spacetime and their use for interstellar travel: A tool for teaching general relativity},
\href{https://doi.org/10.1119/1.15620}{Am. J. Phys., \textbf{56} (1988) 395}.

\bibitem{TW5}
M. S. Morris, K. S. Thorne, and U. Yurtsever, \textit{Wormholes, time machines, and the weak energy condition}, \href{https://doi.org/10.1103/PhysRevLett.61.1446}{Phys. Rev. Lett., \textbf{61} (1988) 1446}.


\bibitem{Falco1}
V. De Falco, E. Battista, S. Capozziello \textit{et al.}, \textit{Reconstructing wormhole solutions in curvature based Extended Theories of Gravity}, \href{https://doi.org/10.1140/epjc/s10052-021-08958-4}{Eur. Phys. J. C, \textbf{81} (2021) 157}.

\bibitem{Falco2} 
V. De Falco, E. Battista, S. Capozziello, and M. De Laurentis, \textit{Testing wormhole solutions in extended gravity through the Poynting-Robertson effect}, \href{https://doi.org/10.1103/PhysRevD.103.044007}{Phys. Rev. D, \textbf{103} (2021) 044007}.

\bibitem{Falco3} 
V. De Falco and S. Capozziello, \textit{Static and spherically symmetric wormholes in metric-affine theories of gravity}, \href{https://doi.org/10.1103/PhysRevD.108.104030}{Phys. Rev. D, \textbf{108} (2023) 104030}.

\bibitem{TW6}
S. Rastgoo and F. Parsaei, \textit{Traversable wormholes satisfying energy conditions in $f(Q)$ gravity},
\href{https://doi.org/10.1140/epjc/s10052-024-12939-8}{Eur. Phys. J. C \textbf{84} (2024) 563}.

\bibitem{TW7}
J. L. Rosa, N. Ganiyeva, and F. S. N. Lobo, \textit{Non-exotic traversable wormholes in $f(R,T_{ab}T^{ab})$ gravity}, 
\href{https://doi.org/10.1140/epjc/s10052-023-12232-0}{Eur. Phys. J. C \textbf{83} (2023) 1040}.

\bibitem{TW8}
P. H. R. S. Moraes, A. S. Agrawal and B. Mishra, \textit{Wormholes in the $f(R,L,T)$ theory of gravity},
\href{https://doi.org/10.1016/j.physletb.2024.138818}{Phys. Lett. B \textbf{855} (2024) 138818}.

\bibitem{TW9}
G. Mustafa, Z. Hassan, P. H. R. S. Moraes and P. K. Sahoo, \textit{Wormhole solutions in symmetric teleparallel gravity}, 
\href{https://doi.org/10.1016/j.physletb.2021.136612}{Phys. Lett. B \textbf{821} (2021) 136612}. 


\bibitem{Radhakrishnan2024}
R. Radhakrishnan, P. Brown, J. Matulevich, E. Davis, D. Mirfendereski, G. Cleaver,
\textit{A review of stable, traversable wormholes in f(R) gravity theories},
\href{https://doi.org/10.3390/sym16081007}{Symmetry, \textbf{16} (2024) 1007}.


\bibitem{Kibble}
T. W. B. Kibble, \textit{Lorentz invariance and the gravitational field} \href{https://doi.org/10.1063/1.1703702}{J. Math. Phys. \textbf{2} (1961) 212--221}.

\bibitem{Smolin}
J. Magueijo, L. Smolin,  \textit{Lorentz invariance with an invariant energy scale}, \href{https://doi.org/10.1103/PhysRevLett.88.190403}{Phys. Rev. Lett. \textbf{88} (2002) 190403}.

\bibitem{C-1}
V.A. Kosteleck{\`y}, \textit{Gravity, Lorentz violation, and the standard model}, \href{https://doi.org/10.1103/PhysRevD.69.105009}{Phys. Rev. D \textbf{69} (2004) 105009}.

\bibitem{C-2}
D. Colladay, V.A. Kosteleck{\`y}, \textit{Lorentz-violating extension of the standard model}, \href{https://doi.org/10.1103/PhysRevD.58.116002}{Phys. Rev. D \textbf{58} (1998) 116002}.

\bibitem{C-3}
T. Clifton, P.G. Ferreira, A. Padilla, C. Skordis, \textit{Modified gravity and cosmology}, \href{https://doi.org/10.1016/j.physrep.2012.01.001}{Phys. Rep. \textbf{513} (2012) 1--189}.

\bibitem{C-4}
S.M. Carroll, et. al., \textit{Cosmology of generalized modified gravity models}, \href{https://doi.org/10.1103/PhysRevD.71.063513}{Phys. Rev. D \textbf{71} (2005) 063513}.

\bibitem{C-5}
S. Nojiri, S.D. Odintsov, \textit{Unified cosmic history in modified gravity: from F (R) theory to Lorentz non-invariant models}, \href{https://doi.org/10.1016/j.physrep.2011.04.001}{Phys. Rep. \textbf{505} (2011) 59--144}.

\bibitem{KS1989}
V. A. Kostelecký, S. Samuel,
\textit{Spontaneous breaking of Lorentz symmetry in string theory},
\href{https://doi.org/10.1103/PhysRevD.39.683}{Phys. Rev. D, \textbf{39} (1989) 683}.

\bibitem{ColladayKostelecky1997}
D. Colladay, V. A. Kostelecký,
\textit{CPT violation and the Standard Model},
\href{https://doi.org/10.1103/PhysRevD.55.6760}{Phys. Rev. D, \textbf{55} (1997) 6760}.

\bibitem{23}
R. Bluhm, V.A. Kostelecký, and N. Russell,
\textit{Testing CPT with anomalous magnetic moments},
\href{https://doi.org/10.1103/PhysRevLett.79.1432}{Phys. Rev. Lett. \textbf{79} (1997) 1432}.

\bibitem{24}
R. Bluhm, V.A. Kostelecký, and N. Russell,
\textit{CPT and Lorentz tests in hydrogen and antihydrogen},
\href{https://doi.org/10.1103/PhysRevLett.82.2254}{Phys. Rev. Lett. \textbf{82} (1999) 2254}.

\bibitem{25}
R. Bluhm, V.A. Kostelecký, and C.D. Lane,
\textit{CPT and Lorentz tests with muons},
\href{https://doi.org/10.1103/PhysRevLett.84.1098}{Phys. Rev. Lett. \textbf{84} (2000) 1098}.

\bibitem{26}
R. Bluhm, V.A. Kostelecký, C.D. Lane, and N. Russell,
\textit{Clock-comparison tests of Lorentz and CPT symmetry in space},
\href{https://doi.org/10.1103/PhysRevLett.88.090801}{Phys. Rev. Lett. \textbf{88} (2002) 090801}.

\bibitem{27}
R. Jackiw and V.A. Kostelecký,
\textit{Radiatively induced Lorentz and CPT violation in electrodynamics},
\href{https://doi.org/10.1103/PhysRevLett.82.3572}{Phys. Rev. Lett. \textbf{82} (1999) 3572}.

\bibitem{28}
J.M. Chung and B.K. Chung,
\textit{Induced Lorentz and CPT violating Chern–Simons term in QED: Fock–Schwinger proper time method},
\href{https://doi.org/10.1103/PhysRevD.63.105015}{Phys. Rev. D \textbf{63} (2001) 105015}.

\bibitem{29}
O.A. Battistel and G. Dallabona,
\textit{Role of ambiguities and gauge invariance in the calculation of the radiatively induced Chern–Simons shift in extended QED},
\href{https://doi.org/10.1016/S0550-3213(01)00304-2}{Nucl. Phys. B \textbf{610} (2001) 316}.

\bibitem{30}
A. P. B. Scarpelli, M. Sampaio, M. C. Nemes, and B. Hiller,
\textit{Chiral anomaly and CPT invariance in an implicit momentum space regularization framework},
\href{https://doi.org/10.1103/PhysRevD.64.046013}{Phys. Rev. D \textbf{64} (2001) 046013}.

\bibitem{31}
T. Mariz, et al.,
\textit{A remark on Lorentz violation at finite temperature},
\href{https://doi.org/10.1088/1126-6708/2005/10/019}{JHEP \textbf{10} (2005) 019}.

\bibitem{32}
J.R. Nascimento, E. Passos, A.Yu. Petrov, and F.A. Brito,
\textit{Lorentz–CPT violation, radiative corrections and finite temperature},
\href{https://doi.org/10.1088/1126-6708/2007/06/016}{JHEP \textbf{06} (2007) 016}.

\bibitem{33}
A.P.B. Scarpelli, M. Sampaio, M.C. Nemes, and B. Hiller,
\textit{Gauge invariance and the CPT and Lorentz violating induced Chern–Simons-like term in extended QED},
\href{https://doi.org/10.1140/epjc/s10052-008-0677-4}{Eur. Phys. J. C \textbf{56} (2008) 571}.

\bibitem{34}
O.M. Del Cima, J.M. Fonseca, D.H.T. Franco, and O. Piguet,
\textit{Lorentz and CPT violation in QED revisited: a missing analysis},
\href{https://doi.org/10.1016/j.physletb.2010.03.081}{Phys. Lett. B \textbf{688} (2010) 258}.

\bibitem{35}
S.M. Carroll, G.B. Field, and R. Jackiw,
\textit{Limits on a Lorentz- and parity-violating modification of electrodynamics},
\href{https://doi.org/10.1103/PhysRevD.41.1231}{Phys. Rev. D \textbf{41} (1990) 1231}.

\bibitem{36}
A.A. Andrianov, R. Soldati, and L. Sorbo,
\textit{Dynamical Lorentz symmetry breaking from a (3+1)-dimensional axion–Wess–Zumino model},
\href{https://doi.org/10.1103/PhysRevD.59.025002}{Phys. Rev. D \textbf{59} (1998) 025002}.

\bibitem{37}
R. Lehnert and R. Potting,
\textit{Vacuum Čerenkov radiation},
\href{https://doi.org/10.1103/PhysRevLett.93.110402}{Phys. Rev. Lett. \textbf{93} (2004) 110402}.

\bibitem{38}
C. Kaufhold and F.R. Klinkhamer,
\textit{Vacuum Cherenkov radiation and photon triple-splitting in a Lorentz-noninvariant extension of quantum electrodynamics},
\href{https://doi.org/10.1016/j.nuclphysb.2005.11.001}{Nucl. Phys. B \textbf{734} (2006) 1}.

\bibitem{39}
H. Belich, L.D. Bernald, P. Gaete, and J.A. Helayël-Neto,
\textit{The photino sector and a confining potential in a supersymmetric Lorentz-symmetry-violating model},
\href{https://doi.org/10.1140/epjc/s10052-013-2632-2}{Eur. Phys. J. C \textbf{73} (2013) 2632}.

\bibitem{40}
G. Gazzola, et al.,
\textit{QED with minimal and nonminimal couplings: on the quantum generation of Lorentz-violating terms in the pure photon sector},
\href{https://doi.org/10.1088/0954-3899/39/3/035002}{J. Phys. G: Nucl. Part. Phys. \textbf{39} (2012) 035002}.

\bibitem{41}
A.P. Baêta Scarpelli,
\textit{QED with chiral nonminimal coupling: aspects of the Lorentz-violating quantum corrections},
\href{https://doi.org/10.1088/0954-3899/39/12/125001}{J. Phys. G: Nucl. Part. Phys. \textbf{39} (2012) 125001}.
\bibitem{42}
B. Agostini, F. A. Barone, F. E. Barone, P. Gaete, J. A. Helay{\"e}l-Neto,
\textit{Consequences of vacuum polarization on electromagnetic waves in a Lorentz-symmetry breaking scenario},
\href{https://doi.org/10.1016/j.physletb.2012.01.050}{Phys. Lett. B \textbf{708} (2012) 212}. 

\bibitem{43}
L. C. T. Brito, H. G. Fargnoli, and A. P. Baêta Scarpelli, \textit{Aspects of quantum corrections in a Lorentz-violating extension of the Abelian Higgs model},
\href{https://doi.org/10.1103/PhysRevD.87.125023}{Phys. Rev. D \textbf{87} (2013) 125023}.

\bibitem{44}
D. Colladay and V.A. Kostelecký,
\textit{Cross sections and Lorentz violation},
\href{https://doi.org/10.1016/S0370-2693(01)00649-9}{Phys. Lett. B \textbf{511} (2001) 209}.

\bibitem{45}
R. Lehnert,
\textit{Dirac theory within the Standard-Model Extension},
\href{https://doi.org/10.1063/1.1769105}{J. Math. Phys. \textbf{45} (2004) 3399}.

\bibitem{46}
B. Altschul,
\textit{Compton scattering in the presence of Lorentz and CPT violation},
\href{https://doi.org/10.1103/PhysRevD.70.056005}{Phys. Rev. D \textbf{70} (2004) 056005}.

\bibitem{47}
G.M. Shore,
\textit{Strong equivalence, Lorentz and CPT violation, anti-hydrogen spectroscopy and gamma-ray burst polarimetry},
\href{https://doi.org/10.1016/j.nuclphysb.2005.03.040}{Nucl. Phys. B \textbf{717} (2005) 86}.

\bibitem{bumblebee}
W. Liu, X. Fang, J. Jing, J. Wang,
\textit{Lorentz violation induces isospectrality breaking in Einstein-bumblebee gravity theory},
\href{https://doi.org/10.1007/s11433-024-2405-y}{Sci. China Phys. Mech. Astron. \textbf{67} (2024) 280413}.

\bibitem{Eslam}
B. Eslam Panah, 
\textit{Super-entropy bumblebee AdS black holes},
\href{https://doi.org/10.1016/j.physletb.2025.139273}{Phys. Lett. B \textbf{861} (2025) 139273}.

\bibitem{Ovgun2019}
A. Övgün, K. Jusufi, İ. Sakallı,
\textit{Exact traversable wormhole solution in bumblebee gravity},
\href{https://doi.org/10.1103/PhysRevD.99.024042}{Phys. Rev. D, \textbf{99} (2019) 024042}.

\bibitem{Oliveira2018}
R. Oliveira, D. M. Dantas, V. Santos, C. A. S. Almeida,
\textit{Quasinormal modes of a bumblebee wormhole},
\href{https://doi.org/10.1088/1361-6382/ab1873}{Class. Quantum Grav., \textbf{36} (2018) 105013}.

\bibitem{Ding2024}
C. Ding, S. Chen, J. Wang, J. Jing,
\textit{Phantom hairy black holes and wormholes in Einstein--bumblebee gravity},
\href{https://doi.org/10.1140/epjc/s10052-025-13815-9}{Eur. Phys. J. C, \textbf{84} (2024) 742}.

\bibitem{Maluf2020}
R. V. Maluf, J. C. S. Neves, C. A. S. Oliveira,
\textit{Modified black hole solution with a background Kalb--Ramond field},
\href{https://doi.org/10.1140/epjc/s10052-020-7902-1}{Eur. Phys. J. C, \textbf{80} (2020) 786}.


\bibitem{Ghosh2024}
S. G. Ghosh, S. D. Maharaj, U. Papnoi,
\textit{Electrically charged black holes in gravity with a background Kalb--Ramond field},
\href{https://doi.org/10.1140/epjc/s10052-024-13188-5}{Eur. Phys. J. C, \textbf{84} (2024) 798}.

\bibitem{MalufMuniz2021}
R. V. Maluf, C. R. Muniz,
\textit{Exact solution for a traversable wormhole in a curvature-coupled antisymmetric background field},
\href{https://doi.org/10.1140/epjc/s10052-022-10409-7}{Eur. Phys. J. C, \textbf{82} (2022) 445}.

\bibitem{LV}
R. B. Magalhães, L. A. Lessa, M. M. Ferreira Jr.,
\textit{Wormholes in Lorentz-violating gravity},
\href{https://doi.org/10.48550/arXiv.2505.07590}{arXiv:2505.07590 [gr-qc] (2025)}.

\bibitem{Guvendi-PLB-1}
A. Guvendi, O. Mustafa, S. Gurtas Dogan,
\textit{Coupled fermion-antifermion pairs within a traversable wormhole},
\href{https://doi.org/10.1016/j.physletb.2025.139313}{Phys. Lett. B, \textbf{862} (2025) 139313}.

\bibitem{Guvendi-PLB-2}
A. Guvendi, H. Hassanabadi,
\textit{Fermion-antifermion pair in magnetized optical wormhole background},
\href{https://doi.org/10.1016/j.physletb.2023.138045}{Phys. Lett. B, \textbf{843} (2023) 138045}.

\bibitem{LV-1}
S. Gurtas Dogan, O. Mustafa, A. Karabulut, A. Guvendi,
\textit{Wave Propagation and Effective Refraction in Lorentz-Violating Wormhole Geometries},
\href{https://doi.org/10.48550/arXiv.2602.10889}{arXiv:2602.10889 [gr-qc] (2026}).

\bibitem{abdelghani-2025}
A. Errehymy, A. Guvendi, S. Gurtas Dogan, O. Mustafa,
\textit{Frame-dragging and light deflection in rotating optical wormhole spacetimes},
\href{https://doi.org/10.1016/j.physletb.2025.139847}{Phys. Lett. B, \textbf{869} (2025) 139847}.


\bibitem{o1}
S. Gurtas Dogan, A. Guvendi, and O. Mustafa,
\textit{Geometric and wave optics in a BTZ optical metric-based wormhole},
\href{https://doi.org/10.1016/j.physletb.2025.139824}{Phys. Lett. B \textbf{868} (2025) 139824}.


\bibitem{reff} 
S. Gurtas Dogan, O. Mustafa, G. Turgut, A. Guvendi,
\textit{Ray and wave optics in twisted graphene nanoribbons},
\href{https://doi.org/10.1088/1402-4896/addf19}{Phys. Scr., \textbf{100} (2025) 075512}.

\bibitem{SOA-2025}
S. Gurtas Dogan, O. Mustafa, A. Guvendi,
\textit{Ray and wave optics in Bonnor-Melvin domain walls: Photon rings},
\href{https://doi.org/10.1016/j.nuclphysb.2025.116920}{Nuc. Phys. B \textbf{1016} (2025) 116920}.

\bibitem{AO-Int}
A. Guvendi, O. Mustafa, 
\textit{Photon rings in three-dimensional Bonnor-Melvin magnetic spacetime},
\href{https://doi.org/10.1142/S0219887825502974}{Int. J. Geom. Methods Mod. Phys. (2025)}.

\bibitem{o2} 
S. Gurtas Dogan, A. Guvendi, O. Mustafa,
\textit{Ray and wave optics in an optical wormhole}, 
\href{https://doi.org/10.1016/j.physletb.2025.139626}{Phys. Lett. B \textbf{868} (2025) 139626}.

\bibitem{o3} 
S. Gurtas Dogan, A. Guvendi, O. Mustafa,
\textit{Ray geodesics and wave propagation on the Beltrami surface: optics of an optical wormhole},
\href{https://doi.org/10.1140/epjc/s10052-025-14644-6}{Eur. Phys. J. C, \textbf{85} (2025) 896}.

\bibitem{o4} 
S. Gurtas Dogan, O. Mustafa, A. Guvendi, H. Hassanabadi,
\textit{Optics in spiral dislocation spacetime: Torsion as a geometric waveguide and frequency-filtering mechanism},
\href{https://doi.org/10.1140/epjc/s10052-025-15239-x}{Eur. Phys. J. C, \textbf{86} (2026) 31}.

\bibitem{o5} 
F. Ahmed, A. Bouzenada,
\textit{Ray and wave dynamics in a static FRW spacetime with conical singularity},
\href{https://doi.org/10.1016/j.physletb.2025.139743}{Phys. Lett. B, \textbf{868} (2025) 139743}.

\bibitem{o6} 
F. Ahmed, A. Bouzenada,
\textit{Lorentz-violating and topological effects on gravitational lensing phenomena and wave optics in wormhole backgrounds},
\href{https://doi.org/10.1016/j.dark.2025.102111}{Phys. Dark Univ., \textbf{50} (2025) 102111}.

\bibitem{o7} 
F. Ahmed, A. Bouzenada,
\textit{Ray geodesic and wave propagation in lorentz-violating wormholes with topological defects},
\href{https://doi.org/10.1088/1402-4896/ae26e7}{Phys. Scr., \textbf{100} (2025) 125306}.

\bibitem{TO1} U. Leonhardt, T.G. Philbin, "Transformation optics and the geometry of light" \href{https://doi.org/10.1016/S0079-6638(08)00202-3}{Progress in optics, \textbf{53} (2009) 69--152}


\bibitem{TO2} D.A. Genov, S. Zhang, X. Zhang, "Mimicking celestial mechanics in metamaterials" \href{https://doi.org/10.1038/nphys1338}{Nature Physics, \textbf{5} (2009) 687--692}


\bibitem{TO3} J. Dalibard, F. Gerbier, G. Juzeli{\=u}nas, P. {\"O}hberg, "Colloquium: Artificial gauge potentials for neutral atoms" \href{https://doi.org/10.1103/RevModPhys.83.1523}{Reviews of Modern Physics, \textbf{83} (2011) 1523--1543}


\bibitem{TO4} M.C. Rechtsman, J.M. Zeuner, Y. Plotnik, Y. Lumer, D. Podolsky, F. Dreisow, S. Nolte, M. Segev, A. Szameit, "Photonic Floquet topological insulators" \href{https://doi.org/10.1038/nature12066}{Nature, \textbf{496} (2013) 196--200}

\bibitem{Konoplya} 
R. A. Konoplya, A. Zhidenko,
\textit{Passage of radiation through wormholes of arbitrary shape},
\href{https://doi.org/10.1103/PhysRevD.81.124036}{Phys. Rev. D, \textbf{81} (2010) 124036}.

\bibitem{Ahmed} 
M. G. Kurbah, F. Ahmed,
\textit{Electromagnetic fields in topologically charged traversable wormholes},
\href{https://doi.org/10.1140/epjc/s10052-024-13566-z}{Eur. Phys. J. C, \textbf{84} (2024) 1205}.

\bibitem{KG} 
R. A. Konoplya, A. Zhidenko,
\textit{Quasinormal modes of black holes: From astrophysics to string theory},
\href{https://doi.org/10.1103/RevModPhys.83.793}{Rev. Mod. Phys., \textbf{83} (2011) 793--836}.



}


\end{thebibliography}
\end{document}